\documentclass[12pt,preprint]{aastex}

\slugcomment{to appear in ApJ Letters}

\shorttitle{Observing Vela X PWN with {\it Fermi LAT}}
\shortauthors{de Jager, Slane \& laMassa}

\begin{document}


\title{Probing the radio to X-ray connection of the Vela X PWN with {\it Fermi LAT} and {\it H.E.S.S.}}


\author{O.C. de Jager\altaffilmark{1}}
\affil{Unit for Space Physics, North-West University, Potchefstroom 2520, South Africa}
\altaffiltext{1}{South African Department of Science \& Technology and National Research Foundation
Research Chair: Astrophysics \& Space Science}
\author{P.O. Slane}
\affil{Harvard-Smithsonian Center for Astrophysics, 60 Garden Street, Cambridge, MA 02138}
\author{S. LaMassa}
\affil{The Johns Hopkins University, 366 Bloomberg Center, 3400 N. Charles Street, Baltimore, MD21218 }


\begin{abstract}
Morphologically it appears as if the \objectname{Vela X} PWN consists of two emission regions:
whereas X-ray ($\sim 1$ keV) and very high energy (VHE) {\it H.E.S.S.} 
$\gamma$-ray observations appear to define a cocoon type shape south of the pulsar,
radio observations reveal an extended area of size $2^{\circ}\times 3^{\circ}$ (including
the cocoon area), also south of the Vela pulsar. Since no wide field of view (FoV) observations 
of the synchrotron emission between radio and X-rays are available, we do not know how the
lepton ($e^\pm$) spectra of these two components connect and how the morphology changes with energy.
Currently we find that two distinct lepton spectra describe the respective radio 
and X-ray/VHE $\gamma$-ray spectra, with a field strength of $5\mu$G self-consistently 
describing a radiation spectral break (or energy maximum) in the multi-TeV domain
as observed by {\it H.E.S.S.} (if interpreted as IC radiation), 
while predicting the total hard X-ray flux above 20 keV 
(measured by the wide FoV {\it INTEGRAL} instrument) within a factor of two. If this same field strength
is also representative of the radio structure (including filaments),
the implied IC component corresponding to the highest radio frequencies
should reveal a relatively bright high energy $\gamma$-ray structure and {\it Fermi LAT} should be able to resolve it.
A higher field strength in the filaments would however imply fewer leptons in \objectname{Vela X} and
hence a fainter {\it Fermi LAT} signal.

\end{abstract}




\section{Introduction}
\label{intro}
The Vela-Puppis region was first identified as a bright radio region by \citet{mss55}.
\citet{r58} compared \objectname{Vela X}, Y and Z (the radio counterparts of the Vela SNR) and 
remarked that \objectname{Vela X} is the most prominent radio feature, with the specific distinction
of having an almost flat radio spectrum. 
A total \objectname{Vela X} radio energy spectrum of the form $F_{\nu}\propto \nu^{-0.39\pm 0.03}$ was
measured by \citet{a01}, which was based on flux measurements up to 8.4 GHz. 
However, between 8.4 and 30 GHz, \citet{h04} resolved the filamentary and PWN structures
and found that the spectra are overall harder: Whereas the PWN spectrum is
$F_{\nu}\propto \nu^{0.1 \pm 0.06}$, the filamentary spectrum is of the form
$F_{\nu}\propto \nu^{-0.28 \pm 0.09}$. 

Following {\it EGRET} observations of the Vela region, \citet{d96} showed that the observation
of inverse Compton (IC) scattered HE $\gamma$-rays from the electron component
producing this radio component, should allow us to constrain the maximum radio frequency 
and associated field strength. This electron component should IC scatter three target photon
components: the Cosmic Microwave Background Radiation (CMBR) field, the Galactic far infrared radiation 
(FIR) field as a result of reradiation of dust grains, and the local starlight field \citep{P07}.

Further insight into the nature of \objectname{Vela X} was revealed by the detection of a cocoon of X-ray
emission south of the Vela pulsar by {\it ROSAT} \citep{mo95}. The X-ray data (with limited energy extent
up to 2.4 keV) was shown to be consistent with a thermal origin, but recently \citet{s08} used {\it XMM-Newton}
observations of the southern head of this cocoon to show that the addition of a non-thermal
synchrotron component with photon index as steep as $\sim 2.3$ is favoured by the data.
By rescaling the {\it H.E.S.S.} flux relative to the {\it XMM-Newton} FoV, a field strength
of $5\mu$G was found.
\citet{m95} also identified a bright radio filament running along this cocoon and models
for such filaments require a strong thermal component, pulsar-generated
magnetic flux and relativistic particles \citep{r88}.


Conclusive proof for the existence of relativistic particles in the X-ray cocoon was
revealed by {\it H.E.S.S.} observations of the Vela region \citep{a06}. In fact, {\it H.E.S.S.} detected 
a spectral maximum from \objectname{Vela X}, which allows us to probe the low energy spectral component, as well as
the energy at which the spectral turnover occurs. An overlay of the {\it H.E.S.S.} image is shown in
Fig. 8 of \citet{dd08}, showing the overlap of the VHE $\gamma$-ray cocoon and the bright radio filament
referred to by \citet{m95}. Whereas a hadronic interpretation was suggested by \citet{h06},
\citet{d07} and \citet{dd08} reviewed evidence in favour of a leptonic IC origin. Recently \citet{s08}
also found that the thermal particle density at the head of the cocoon where bright VHE $\gamma$-ray
emission was found, is only 0.1 cm$^{-3}$ - a factor 6 lower than the lower limit employed by \citet{h06}.
This places more stringent constraints on the brightness of a hadronic signal from \objectname{Vela X}.

\citet{m05} derived an X-ray photon index of $1.50 \pm 0.02$ within a distance of 0.5 arcmin
from the pulsar. This index represents the uncooled (by synchrotron losses) 
electron spectral index injected into the PWN, since a steepening is observed 
for distances greater than 0.5 arcmin. The corresponding uncooled post shocked electron spectral
index of 2.0 is also consistent with the pre-break spectral index detected by {\it H.E.S.S.},
if the signal from the latter has an inverse Compton origin. 

\citet{d07} has shown that \objectname{Vela X} most likely consists of a 2-component
non-thermal spectrum: From Fig. 2 of \citet{d07} the lepton spectrum corresponding
to the VHE $\gamma$-ray spectrum only connects to the radio lepton spectrum if the 
field strength of the radio structure is $>200 \mu$G, if we extrapolate
the uncooled lepton index of 2.0 towards the radio domain. However, 
the field strength in the radio filaments is more likely to be in the range 10 to 50 $\mu$G
\citep{m95}.


The observation of an X-ray photon index $>2.0$ in the cocoon by \citet{s08} may be due to
synchrotron cooling inside the X-ray band (below 7 keV) and to test this we need hard X-ray
observations with a FoV covering the full size of \objectname{Vela X}.
Only {\it INTEGRAL} (20 - 300 keV) has a large enough FoV to cover the extended \objectname{Vela X}
(and by default cocoon) region. However, only point-like unpulsed emission centered on the pulsar
(consistent with the 12 arcmin PSF) with a photon index of $-1.86\pm 0.03$ 
was detected by {\it INTEGRAL} (L. Kuiper, personal communication, 2007). Thus,
synchrotron burnoff must have terminated most emission above 20 keV at distances $>12$ arcmin from the pulsar to levels consistent with the sensitivity of {\it INTEGRAL}.
This spectrum is consistent with the {\it OSSE} spectrum reported by \citet{d96b}, covering at least the total 
cocoon region. The {\it Beppo}SAX PDS, with a FoV of $1.3^{\circ}$ (FWHM), also observed the Vela pulsar above 20 keV,
covering most of the cocoon region \citep{m05}. Although the observed photon index of the nebula
is slightly steeper ($2.00 \pm 0.05$), the energy flux is comparable to that of {\it INTEGRAL}.

In this paper we construct a time dependent post shocked injection spectrum $Q(E,t)$ (with $E$
the lepton energy). This spectrum consists of two components (as motivated a above): 
$Q_R(E,t)$ as source for the extended radio emission
and $Q_X(E,t)$ as source for the X-ray and VHE $\gamma$-ray emission, confined mostly to the cocoon region. In this paper {\it Fermi LAT} observations should be able to test the spectral connection between $Q_X$ and $Q_R$.

These two distinct source spectra are normalised to the time averaged spindown power, giving conversion efficiencies $\eta_R$ and $\eta_X$ of spindown power to leptons.
The time dependent transport equation with synchrotron losses is then solved to produce the present
time electron spectra $dN_R/dE$ and $dN_X/dE$. The field strength required for the synchrotron loss term is obtained
from {\it H.E.S.S.} observations, if we assume that the VHE $\gamma$-ray signal is the result
of IC scattering. Since {\it H.E.S.S.} measured the spectral energy maximum, we
find the magnetic field strength which reproduces the observed multi-TeV spectral break, so that
$dN_X/dE$ is uniquely determined in the energy range 5 to 100 TeV.
We then show that this same $dN_X/dE$ (with $B$ derived from the VHE spectral break) also predicts 
the total hard X-ray flux from the PWN measured by {\it INTEGRAL} correctly within a factor two. 
For $\eta_R<1$ we also show that $dN_R/dE$ (with the abovementioned $B$) reproduces the total integrated radio flux, which in turn predicts a high energy $\gamma$-ray component, which can be tested with {\it Fermi LAT}.

\section{The double source spectrum of Vela X}
\label{2source}
In the discussion below we will neglect (to a first order) the effect of adiabatic cooling on the spectrum \citep{ps73} since the
reverse shock (applicable to \objectname{Vela X}) is expected to have adiabatically heated the particles during the PWN crushing phase. Energy losses due to adiabatic cooling during the pre-crushing phase is then partially cancelled
during the crushing phase (Ferreira \& de Jager 2008, in preparation).

Let $Q_R(E,t)$ and $Q_X(E,t)$ be the respective source spectra corresponding to the radio and X-ray synchrotron components, and $L(t)$ be the time dependent spindown power, of which fractions $\eta_R$ and $\eta_X$ are converted to these two components. Based on the discussion in Section 1, we model these source spectra as
\begin{eqnarray}
Q_R(E,t)=Q_{0,R}(t)(E/E_{{\rm max},R})^{-p}\\
Q_X(E,t)=Q_{0,X}(t)(E/E_{{\rm max},X})^{-2},
\end{eqnarray}
for $E_{{\rm min},R}<E<E_{{\rm max},R}$ and $E_{{\rm min},X}<E<E_{{\rm max},X}$ in the two respective equations. These
upper and lower limits are currently weakly constrained by observations.
The radio spectral index of 0.39 requires a particle index of $p=0.39\times 2+1=1.78$,
whereas an X-ray (VHE $\gamma$-ray) lepton spectral index of 2 would reproduce the uncooled photon spectral index of 1.5.
By requiring the total energy in each respective component to be a fraction $\eta_R<1$ and $\eta_X<1$ 
(with $\eta_R+\eta_X\le 1$)
of the spindown power, we can show that (with the minimum electron energy for the radio component much smaller than the corresponding maximum)
\begin{eqnarray}
Q_R(E,t)\sim \left(\frac{\eta_R L(t)(2-p)}{E_{{\rm max},R}^{2-p}}\right)E^{-p} \\
Q_X(E,t)=\left(\frac{\eta_X L(t)}{\ln\left({E_{{\rm max},X}/E_{{\rm min},X}}\right)}\right)E^{-2}.
\end{eqnarray}
Neglecting adiabatic losses (as discussed above), but including radiation losses with timescale
$\tau_{\rm rad}(E)=E/(dE/dt)_{\rm rad}$, where $(dE/dt)_{\rm rad}$ is the sum of the
synchrotron and IC (Klein-Nishina effects included) energy loss rates,
the transport equation can be solved to give the present day (age $T$) total particle
spectrum for each component $i=R,X$ as (see e.g. \citet{z08})
\begin{equation}
 \frac{dN_i(E,T)}{dE}=\int_0^T Q_i(E,t)\exp\left(-\frac{T-t}{\tau_{\rm rad}(E)}\right)dt.
\label{nx}
\end{equation}
This expression effectively allows leptons to be accumulated over time $T$ in the nebula.
The resulting spectrum $dN/dE$ is however modulated (steepened) at the high energy tail by radiation losses. 
 
Since radiation losses are negligible for the radio component, the integral over the source term
$Q_R$ yields the change in rotational kinetic energy ($\Delta{\rm KE}_{\rm rot}=I(\Omega_0^2-\Omega^2)/2$) since birth, as shown by \citet{d08}, so that (for $p<2$)
\begin{equation}
\frac{dN_R(E,T)}{dE}=\frac{\overline{\eta_R}(2-p)\Delta {\rm KE}_{\rm rot}}{E_{{\rm max},R}^{2-p}}E^{-p}.
\label{nr}
\end{equation}
The term $\overline{\eta_R}$ is the time averaged conversion efficiency. \citet{vw01} found
that a pulsar birth period of $P_0=40$ to 50 ms is required to reproduce the observed ratio ($\sim 0.25$) 
between the PWN and SNR radius. The recently measured pulsar braking index $n=1.6\pm 0.1$ \citep{dod07}
implies an age between $T=11$ and 15 kyr given the abovementioned range for $P_0$. This range of
values (or exact choice of braking index) do not change our conclusions significantly.

\section{The particle, synchrotron and IC spectra of the cocoon}
\label{cocoon}
The lepton spectrum for the X-ray/VHE $\gamma$-ray emission region (mostly the cocoon region) derived from Eqn.(\ref{nx}) 
(parameters in Table 1)
is constrained by the following information:
an injection (uncooled) lepton spectral index of 2.0 as discussed in section~\ref{intro},
the IC normalisation of the VHE cocoon spectrum as provided by {\it H.E.S.S.} \citep{a06},
the multi TeV break energy observed by {\it H.E.S.S.}, spindown from $P_0=40$ ms to the current period
of 89 ms within time $T$ ($\sim 11$ to 15 kyr), an implied time averaged 
conversion efficiency of $\overline{\eta_X}=3\times 10^{-3}$, and a
field strength of $5\mu$G to reproduce the observed VHE break energy. 

It is clear that this model
IC spectrum agrees relatively well with {\it H.E.S.S.} measurements (Fig. 1 - 
thick solid line at very high $\gamma$-ray energies). The CMBR mostly contributes
to this spectrum, whereas higher energy contributions from the galactic FIR and
starlight photon fields are suppressed due to Klein Nishina effects. However, in the next section
we will see that the latter two radiation fields contribute significantly
to the high energy IC $\gamma$-ray counterpart of the radio spectrum, where
the IC scattering is in the Thomson limit. 


The lepton spectrum of the cocoon,
if self-consistent, should be able to predict the total X-ray spectrum from
\objectname{Vela X}. From Fig. 1 it is clear that the model synchrotron spectrum shows a radiation break
around 1 keV ($\sim 2$ to $3\times 10^{17}$ Hz) and predicts the 20 - 300 keV
{\it INTEGRAL} flux within a factor 2, although with a steeper spectrum. The latter is expected
to harden if we allow the field strength to increase towards the pulsar in a more sophisticated model
incorporating a spatial dependence as well.

\section{The particle, synchrotron and IC spectra of the total Vela X PWN}
To first order we assumed the same field strength ($5\times 10^{-6}$G) of the entire 
\objectname{Vela X} structure as found for the cocoon \citep{s08}.
This may not be the case, but by comparing {\it Fermi LAT} with high frequency radio observations 
we should be able to measure the field strength in this extended structure if the high frequency (radio) cutoff or turnover can be independently determined. 
In this case synchrotron cooling is negligible and the steady state spectrum is given by Eqn (\ref{nr}).
To reproduce the absolute radio flux (for a particle index of $p=1.78$) we had to assume efficiencies of
$\overline{\eta_R}=0.35$ and 0.45 corresponding to the assumed spectral cutoffs $10^{10}$ Hz and $10^{11}$ Hz respectively. The leptonic parameters corresponding to the latter cutoff
is listed in Table 1 under ``radio''. 
From Fig. 1 it is clear that the maximum frequency should be $>10^{10}$ Hz.

While the lack of high-frequency radio measurements prohibits a clear
constraint on the cutoff of this component, {\it Fermi LAT} observations can
constrain the maximum of the spectral energy distribution 
of leptons in the radio nebula. As an example
we employ cutoffs at the abovementioned two frequencies to calculate the corresponding 
expected high energy $\gamma$-ray spectra. These predictions are 
shown as dashed lines in Fig. 1 (marked ``HE $\gamma$-ray IC'').
The hard lepton spectrum followed by an assumed cutoff at an electron energy $E_{{\rm max},R}=\gamma_{{\rm max},R}mc^2$
should produce broad $\gamma$-ray features at three energies: $3.6\gamma_{{\rm max},R}^2kT_i$
corresponding to the temperatures $T_i=2.7$K (CMBR), 25K (FIR) and $\sim 4000$K to $\sim 8000$K (starlight).

Although these energy densities are comparable in the local neighbourhood \citep{P07}, we
investigate deviations at the location of Vela X: The \objectname{IRAS-Vela shell} (IVS)
of dark clouds and cometary globules centered on 
($\ell^{II}=259.85^{\circ}$, $b^{II}=-8.25^{\circ}$) with a radius of $(5.1^{\circ}\pm 0.2^{\circ})$
and annular RMS width of $\sim 2^{\circ}$ is located at a distance $\sim 400$ pc \citep{t06}. With a $100\mu$m IRAS ring flux of $\sim 10$MJy/sr we obtain a luminosity of $\sim 10^{38}$ erg/s, and 
assuming a distance of $r=100$ pc between the IVS center and Vela X, we obtain an additional 
FIR energy density of $0.002(100\;{\rm pc}/r)^2$ eV/cm$^3$ at Vela X,
which is well below our assumed local averaged energy density. Other potential target photon fields from
the Vela molecular ridge and Gum Nebula are located directly behind the Vela pulsar \citep{Sahu92}
so that the contribution from these local clouds would be severely suppressed by the head-on-tail
effect of the IC process, since the $(1-\cos\theta)^2$ term (with $\theta$ small) in the energy
loss rate term is much less than unity.

Furthermore, fluctuations in the local distribution of stars can also produce deviations in the starlight
energy density at Vela X. To investigate this we extracted the {\it Hipparcos} catalog of $\sim 118,000$ stars,
of which $\sim 73,000$ have parallax measurements at the $>3\sigma$ detection level. This subset was used to calculate
the starlight energy density at any 3D field point within $\sim 300$ pc from Earth. Within $\sim 20$ pc from Earth
we obtain a value of $U_{\rm rad}=0.24$ eV/cm$^3$, consistent with \citet{P07}, 
whereas a value of $U_{\rm rad}=0.44\pm 0.12$ eV/cm$^3$ is obtained for the Vela PWN,
with the error reflecting the uncertainty in the distance to the Vela pulsar \citep{dod03}.  
This mean value (0.44 eV/cm$^3$) was used in Fig. 1, but
assuming a mean temperature of 7500 K to account for the stellar types contributing
to this excess. 

Whereas the {\it Fermi LAT} webpage \footnote{$http://www-glast.slac.stanford.edu/software/IS/glast\_lat\_performance.htm$}
only lists sensitivities for point sources, we scale
the actual {\it EGRET} upper limits for the 
Vela X extended source (extracted from the unpulsed skymaps) obtained by \citet{d96} by a conservative 
factor of 30 downward to account for the improved {\it Fermi LAT} sensitivity in the same energy bands
for an integration time of 1 year.
The {\it EGRET} integral flux upper limits and {\it GLAST/Fermi LAT} sensitivities are shown in Fig. 1.

\section{Discussion and Conclusion}
In this paper we have shown that a pulsar birth period of $\sim 40$ ms (to reproduce the observed ratio
of PWN to SNR radii), spinning down within 11 to 15 kyr to the present period of 89 ms while 
converting $\eta_X\sim 0.3\%$ of its spindown power to leptons (e$^{\pm}$)
in the PWN with a spectral index of 2, would reproduce the observed {\it H.E.S.S.}
flux in the \objectname{Vela X} cocoon via the IC process. 
A nebular averaged field strength of $5\mu$G would also reproduce the observed TeV spectral break, 
Furthermore, this same spectrum
predicts the {\it Beppo}SAX PDS and {\it INTEGRAL} total fluxes within a factor of two. 

Assuming the same 5$\mu$G field strength in the much larger $2^{\circ}\times 3^{\circ}$ \objectname{Vela X} radio nebula,
we need to invoke a second e$^{\pm}$ component with electron spectral index 1.78 and a much larger conversion
efficiency $\eta_R\sim 0.4$ 
to reproduce the observed total radio spectrum up to at least 8.9 MHz and possibly 30 MHz. This spectrum 
should cut off or steepen significantly to meet the less energetic cocoon spectrum.

The implication of this second (radio) component is the prediction of an IC component at HE
$\gamma$-rays covering the same $2^{\circ}\times 3^{\circ}$ area south of the pulsar. 
The visibility of this
component in the HE $\gamma$-ray domain for {\it Fermi LAT} is further improved due to the 
existence of the local FIR and starlight target photon fields, scattered by the highest energy leptons
in this second component up to $\sim 10 $ GeV $\gamma$-rays (in the Thomson limit) 
where {\it Fermi LAT} has good sensitivity. 


If the electrons radiating the radio component  are trapped in compressed regions (e.g. filaments) 
with field strength $\gg 5\mu$G, we would require fewer leptons and hence lower efficiency $\eta_R$ 
to reproduce the radio emission. A HE $\gamma$-ray flux measurement, or upper limit thereof, will
provide constraints on this field and hence $\eta_R$. 

\acknowledgments
PO Slane acknowledges support from NASA contract NAS8-39073. The referee is thanked for useful comments.

\clearpage

\begin{table}
\begin{center}
\caption{Parameters associated with the two lepton components of Vela X for 
a pulsar birth period of 40 ms, and an assumed braking index of 2.5 to
give an age of 11 kyr. $E_{\rm tot}$ is the total lepton energy between
$E_{\rm min}$ and $E_{\rm max}$.}
\tablewidth{0pt}
\begin{tabular}{lcccccc}
\tableline
& $\eta$ & $p$ & $E_{\rm min}$ & $E_{\rm max}$ & $\nu_{\rm max}$ & $E_{\rm tot}$ \\
&        &     &               &    (eV)       &      (Hz)      &   (erg)  \\
\tableline
radio & 0.35   & 1.78 & $m_ec^2$ & $2\times 10^{10}$ & $10^{10}$ & $3\times 10^{48}$ \\
radio & 0.45   & 1.78  & $m_ec^2$ & $6\times 10^{10}$ & $10^{11}$ & $4\times 10^{48}$ \\
X-ray & 0.003 & 2.0  & $m_ec^2$ & $3\times 10^{15}$ & $2\times 10^{20}$ & $2\times 10^{46}$ \\
\tableline
\end{tabular}
\end{center}
\end{table}



\clearpage

\begin{figure}
\plotone{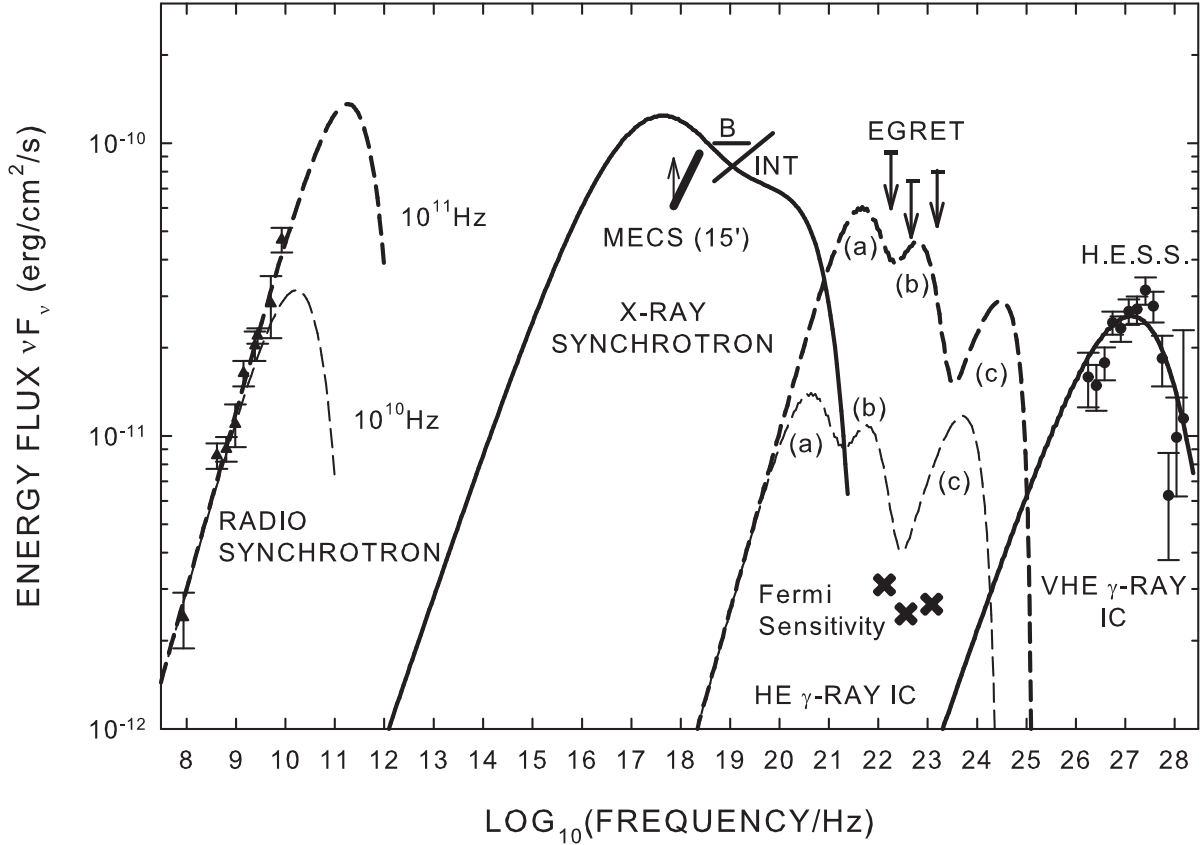}
\caption[]{{\bf Dashed lines}: The radio synchrotron and high energy (HE) IC $\gamma$-ray
spectra of the total $2^{\circ}\times 3^{\circ}$ \objectname{Vela X} structure
corresponding to synchrotron cutoffs of $10^{10}$ Hz ({\it thick dashed line}) and $10^{11}$ Hz
({\it thin dashed line}). HE IC $\gamma$-ray features from scattering on {\bf (a)} the CMBR
{\bf (b)} FIR and {\bf (c)} starlight.
{\bf Thick solid lines}: The X-ray synchrotron and VHE $\gamma$-ray IC spectra including
spin evolution. {\bf Triangles:} Radio flux measurements from
\citet{a01}. {\bf Circles:} {\it H.E.S.S.} data from \citet{a06}.
{\bf Short thick solid lines:} {\it INTEGRAL} (``INT'') data from
L. Kuiper (2007, personal communication), 
{\it Beppo}SAX ``MECS'' shown as a lower limit for an extraction radius of 15 arcmin and
PDS (``B'') from \citet{m05}. {\bf Arrows}: {\it EGRET} Vela X ULs from \citet{d96}.
{\bf Crosses}: {\it Fermi LAT} 1-year sensitivity from EGRET ULs, scaled down by $\times 30$.}

\end{figure}



\begin{thebibliography}{}

\bibitem[Aharonian et al. (2006)]{a06} Aharonian, F.A., et al. (H.E.S.S. Collaboration), 2006, A\&A, 448, L43
\bibitem[Alvarez et al.(2001)]{a01} Alvarez, H., Aparici, J., May, J., \& Reich, P. 2001, A\&A, 372, 636
\bibitem[de Jager et al.(1996a)]{d96} de Jager, O.C., Harding, A.K., Sreekumar, P., \& Strickman, M. 1996, 
A\&AS., 120, 441
\bibitem[de Jager et al.(1996b)]{d96b} de Jager, O.C., Harding, A.K., Strickman, M.S. 1996, \apj, 460, 729
\bibitem[de Jager (2007)]{d07} de Jager, O.C. 2007, \apj, 658, 1177
\bibitem[de Jager (2008)]{d08} de Jager, O.C. 2008, \apj, 678, L113
\bibitem[de Jager \& Djannati-Ata\"i(2008)]{dd08} de Jager, O.C. \& Djannati-Ata\"i, A. 2008,
to appear in Springer Lecture Notes on Neutron Stars and Pulsars: 40 years after their discovery, eds. W. Becker,	arXiv:0803.0116v1 (astro-ph)
	
\bibitem[Dodson et al.(2003)]{dod03}Dodson, R., Legge, D., Reynolds, J. E. \& McCulloch, P. M. 2003,
\apj, 596, 1137
\bibitem[Dodson, Lewis \& McCulloch(2007) ]{dod07} Dodson, R, Lewis, D. \& McCulloch, P. 2007, 
Ap\&SS, 308, 585
\bibitem[Hales et al.(2004)]{h04} Hales, A.S., Casassus, S., Alvarez, H., May, J., Bronfman, L., Readhead, A.C.,
Pearson, T.J., Mason, B.S. \& Dodson, R., \apj, 613, 977
\bibitem[Horns, Aharonian \& Santangelo(2006)]{h06} Horns, D., Aharonian, F.A., \& Santangelo, A., et al.~ 2006, A\&A, 451, L51
\bibitem[LaMassa, Slane \& de Jager (2008)]{s08} LaMassa, S., Slane, P.O. \& de Jager, O.C. 2008, submitted to \apj
\bibitem[Mangano et al.(2005)]{m05} Mangano, V., Massaro, E., Bocchino, F.,  Mineo, T. \& Cusumano, G. 2005, A\&A, 436, 917
\bibitem[Markwardt \& \"Ogelman(1995)]{mo95} Markwardt, C.B. \& \"Ogelman, H. 1995, Nature, 375, 40 
\bibitem[McGee, Slee \& Stanley(1955)]{mss55}McGee, R.X., Slee, O.B., \& Stanley, G.J. 1955, Aust. J. Phys., 8, 347
\bibitem[Milne (1995)]{m95} Milne, D.K. 1995, \mnras, 277, 1435
\bibitem[Pacini \& Salvati(1973)]{ps73} Pacini, F. \& Salvati, M. 1973, \apj, 186, 249
\bibitem[Porter et al.(2007)]{P07} Porter, T.A., Digel, S.W., Grenier, I.A., Moskalenko, I.V., \& Strong, A.W. 2007,
in Proc. 30th ICRC, Merida, Mexico, arXiv:0706.0221 (astro-ph)
\bibitem[Reynolds (1988)]{r88} Reynolds, S.P 1988, \apj, 327, 853
\bibitem[Rishbeth (1958)]{r58} Rishbeth, H. 1958, Aust. J. of Phys., 11, 550
\bibitem[Sahu (1992)]{Sahu92} Sahu, M.S. 1992, Ph.D Thesis, Univ. of Groningen 
\bibitem[Testori et al(2006)]{t06} Testori, J.C., Arnal, E.M., Morras, R., Bajaja, E., 
P\"{o}ppel, W.G.L., \& Reich, P. 2006, A\&A, 458, 163
\bibitem[van der Swaluw \& Wu(2001)]{vw01} van der Swaluw, E. \& Wu, Y. 2001, \apj, 555, L49
\bibitem[Zhang, Chen \& Fang (2008)]{z08} Zhang, L., Chen, S.B., \& Fang, J. 2008, \apj, 676, 1210

\end{thebibliography}
\end{document}